\documentclass[12pt]{iopart}
\usepackage{iopams, graphicx}
\usepackage{color}
\definecolor{Red}{rgb}{1,0,0}

\newcommand{\up}{\uparrow}

\newcommand{\Up}{\Uparrow}

\newcommand{\be}{\begin{equation}}
\newcommand{\ee}{\end{equation}}
\newcommand{\beq}{\begin{eqnarray}}
\newcommand{\eeq}{\end{eqnarray}}
\newcommand{\dg}{\dagger}
\newcommand{\zb}{\overline{z}}
\newcommand{\nn}{\nonumber}

\begin{document}

% ------------------------------------------------------------------------------------------------

\title{Charged spin textures over the Moore-Read quantum Hall state}

\author{J Romers$^1$, L Huijse$^2$ and K Schoutens$^1$}
\address{$^1$ Institute for Theoretical Physics, University of Amsterdam,
Science Park 904, P.O.Box 94485, 1090 GL Amsterdam, The Netherlands}
\ead{j.c.romers@uva.nl}
\address{$^2$ Department of Physics, Harvard University, Cambridge MA 02138, USA}

\date{January 15, 2011}

\begin{abstract}
We study charged spin textures (CST) over the Moore-Read quantum Hall
state at filling factor 5/2. We develop an algebraic framework and show that the 
pairing condition that is inherent in the Moore-Read state naturally leads to a
class of charged spin textures, labeled by winding numbers $[w_{I},w_{I\! I}]$.
The fundamental CST, with labels $[1,0]$ and electric charge $e/4$, is identified 
with the polar core vortex known in the spin-1 BEC literature. The spin texture 
carried by the fusion product of fundamental CSTs is correlated with the
fusion channel of the underlying non-Abelian quasiholes.
\end{abstract}


% PACS ok?

\maketitle
% intro-1 spin in fqH states, skyrmion wf, texture, topo charge.
Fractional quantum Hall (fqH) systems are the best known realizations of topological order 
in nature. Among all observed fqH states, the one at $\nu = \frac{5}{2}$ is highly special. 
Its proposed theoretical descriptions, the Pfaffian (Moore-Read) \cite{MR} and anti-Pfaffian 
\cite{aPf} states in particular, are based on a pairing mechanism and, as a consequence, 
admit excitations  with non-Abelian braid statistics.

In fqH systems the magnetic and the
Coulomb energies are of the same order and it is not a priori clear whether or not
the electron spin plays a part in understanding these phases of matter. 
This leaves several possibilities. The first is that the ground state itself can be 
non-polarized. In a second possible scenario the ground state is polarized, but the excitations involve overturned spins. In this paper we analyze spin-full excitations over the MR state, 
which by itself is fully polarized. We analyze {\it charged spin textures} (CST) carrying the
fundamental fractional charge $q=e/4$ and study the spin textures that arise upon
fusing these excitations \footnote{While our results are 
specific to the MR state, we 
expect that the essential insights carry over to a broader class of paired wave functions, 
such as the anti-Pfaffian state.}.

Experiments have confirmed that the lowest energy excitations in $\nu = 1$ integer quantum 
Hall (iqH) systems are charged spin textures. They are well 
described by wave functions of the form \cite{MFB}
\be
\label{iqh}
\psi^{(\lambda)}_{\rm skyrmion} = \psi^{(\lambda)}_{\rm B} \psi_{\rm iqH}, \nn
\ee
where $\psi_{\rm iqH}$ is the ground state wave function for $\nu = 1$ and 
$\psi^{(\lambda)}_{ \rm B}$ is a wave function for bosons. These excitations are
referred to as skyrmions because in an effective field theory approach,
they correspond to field configurations carrying
non-trivial topological charge. This is measured by the Pontryagin
index \cite{Pont}, which is identified with the electric charge carried by the skyrmion.
A typical spin profile (Pontryagin index $+1$, charge $+e$ skyrmion) reads
\be
{\bf S} (r,\phi)= (\sqrt{1-\sigma^2} \cos \phi,-\sqrt{1-\sigma^2} \sin \phi, \sigma) ,
\label{eq:iqHtexture}
\ee
where $\sigma(r)$ is $-1$ at the origin and $+1$ at infinity.

% intro-2:

The recent paper \cite{WMSC}, see also \cite{FRYND,DHN}, 
took up the study of spin-full excitations over the MR state, stressing that the 
results may shed light on some of the experimental findings regarding the 
spin-polarization at $\nu=5/2$. A particular suggestion is that specific experimental 
probes aimed at detecting a spin-polarization at $\nu=5/2$ \cite{SPUMPB} may 
excite spin-full excitations, thereby depolarizing the system. The authors of 
\cite{WMSC} report an extensive numerical study of up to 
$N=20$ particles in spherical geometry, where angular momentum ($L$) and total 
spin ($S$) are good quantum numbers. They
identified low-lying states on the diagonal $L=S$ as well as spin-full ground
states with $L=0$ or $L=1$. The $L=S$ states are associated with charge $2q=e/2$ 
skyrmions. The $L=S=0$ state in particular is well described by a product state
of the form
$
\psi_{\rm skyrmion} = \psi^{(L=S=0)}_{\rm B} \psi_{\rm MR} .
$
The spin-full states with $L=0(1)$ are naturally interpreted as being built from spatially
separated charge $q$ CSTs. The paper \cite{WMSC} presents a phase diagram 
specifying the nature of the spin-full excitations
(skyrmions vs. separated CSTs) as a function of the Zeeman splitting and the lateral harmonic
confinement strength, which is used to model disorder.

% algebraic approach - I

Here we follow a purely algebraic approach to obtain explicit expression for a variety
of CSTs over the MR state. It rests on two observations. The first \cite{CGT} is that the 
MR wavefunction, which is most commonly written in terms of a Pfaffian determinant 
\cite{MR}, can be  expressed as (we focus on the bosonic version with filling $\nu=1$
and fundamental charge $q=e/2$, its fermionic counterpart with $\nu=1/2$
and $q=e/4$ can be obtained by including an overall Jastrow factor)
\be 
\psi_{\rm MR} (\{z_i\}) =
\begin{array}[t]{c} {\rm Symm} \\ {\small I,I\! I} \end{array} 
   \left[ \psi^L_{I} \psi^L_{I\! I} \right]\ . 
\label{symI_II}
\ee 
Here we divide the particles into two groups $I,\ I\!I$ of equal size, write
a bosonic Laughlin wave function 
$ 
\psi^L_{I,I\!I} (\{z_j\})=
\prod_{i<j \in I,I\!I} (z_i-z_j)^2 
$
for each group and then symmetrize over all ways to distribute
the particles over groups $I$ and $I\!I$. The second observation \cite{MFB} is that 
in the lowest Landau Level (LLL), the most general wavefunction for $N$ spin-full 
fermions in $N+1$ available 1-particle orbitals that vanishes when two particles are placed
at the same position \footnote{The Pauli principle only dictates that the wave function should vanish
when two particles of identical spin are placed at the same position. The restriction to states 
that have the additional property that they vanish when two particles of opposite spin are placed at the same
position makes the whole construction far more transparent. The price to pay is that not all possible states are 
reachable by this construction.} factorizes as $\psi_{\rm B} \psi_{\rm iqH}$,
where $\psi_{\rm B}$ describes $N$ spin-1/2 bosons in two orbitals.
These orbitals can be viewed as the 
$L_z=\pm 1/2$ components of a $L=1/2$ doublet of angular momentum.
The combined orbital and spin angular momenta give rise to an
$SU(4)$ symmetry, with the four 1-particle states $(L=S=1/2)$ corresponding to
the fundamental (vector) representation, with Dynkin labels $[1\,0\,0]$ \footnote{For an excellent review on the 
theory of Lie algebras, see \cite{Slansky}. This work also contains lists of irreducible representations
of Lie algebras that arise in physical systems.}. The partition 
sum for $N$ such bosons becomes
\beq 
Z_B &=& SU(4)\ {\rm irrep}\ [N \, 0 \, 0] \nonumber\\
    &=&  \sum_{K=0}^{N/2} \left( L=N/2-K,S=N/2-K \right) \ . 
\eeq
For large $N$ the highest weight states (HW) of each of the $(L=S=N/2-K)$
multiplets simplify to become
\be 
\label{Kturned}
| \psi_{\rm B}^K \rangle 
= {\rm HW} \left(L=S=N/2-K \right)
\rightarrow | \! \downarrow_{K},\uparrow_{N-K}\rangle, \nn
\ee
with the left (right) position in the ket corresponding to the
$L_z=-1/2$ ($+1/2$) orbital. The expression for a size-$\lambda$ skyrmion in the disc geometry is
then obtained as a weighted sum over these states \cite{Moon,MFB},
\be
\left( \sum_K \lambda^K \psi_{\rm B}^K \right) \psi_{\rm iqH} =  \prod_{i<j}
 (z_i-z_j)\prod_l(z_l| \! \uparrow\rangle_l + \lambda | \! \downarrow\rangle_l) \; e^{-\frac{|z_l|^2}{4}}, \nn
\ee
leading to the spin texture eq.~(\ref{eq:iqHtexture}).

% algebraic approach - II

The bosonic MR state is uniquely characterized as the highest density spin-polarized LLL 
state that is annihilated by the pairing hamiltonian
$
H_{\rm pair} = \sum_{i<j<k}\delta(z_i-z_j) \delta(z_j-z_k).
$
We restrict ourselves to spin-full states that satisfy the very same MR pairing condition $H_{\rm pair} \psi =0$. 
For spin-1/2 bosons, the highest density state with this `MR pairing' 
property is the non-Abelian 
spin singlet (NASS) state \cite{NASS} with filling 
fraction $\nu=4/3$. In spherical geometry, the NASS state is realized 
for $N^{\rm NASS}_\phi=3/4 N- 2$ flux quanta. The total space of paired states 
for $N_\phi$ in the vicinity of $N_\phi^{\rm MR} = N-2$ can be understood through 
counting formulas for spin-full quasiholes over the NASS state \cite{NASS}. 
In table \ref{N4Nflux1_2_3} we list the dimensions
of each of the $(L,S)$ subspaces for $N=4$ particles for $N_\phi=N_\phi^{\rm  NASS}=1$, 
$N_\phi=N_\phi^{\rm MR}=2$, $N_\phi=N_\phi^{\rm MR}+1 =3$.

\begin{table}

\begin{tabular}{|c|c|}
\hline
$N_\phi$=1 & $S$=0 \\
\hline
$L$=0 & 1  \\
\hline
\end{tabular}
%\hspace{1mm}
\begin{tabular}{|c|c|c|c|}
\hline
$N_\phi$=2       & $S$=0 & $S$=1 & $S$=2 \\
\hline
$L$=0 & 1 & 0 & 1 \\
\hline
$L$=1 & 0 & 1 & 0 \\
\hline
$L$=2 & 1 & 0 & 0 \\
\hline
\end{tabular}
%\hspace{1mm}
\begin{tabular}{|c|c|c|c|}
\hline
$N_\phi$=3       & $S$=0 & $S$=1 & $S$=2 \\
\hline
$L$=0 & 1 & 0 & 1 \\
\hline
$L$=1 & 0 & 2 & 0 \\
\hline
$L$=2 & 2 & 1 & 1 \\
\hline
$L$=3 & 0 & 1 & 0 \\
\hline
$L$=4 & 1 & 0 & 0 \\
\hline
\end{tabular}

\caption{Multiplicities of $(L,S)$ multiplets in state space for $N=4$ bosonic
spin-1/2 particles on 
the sphere, subjected to MR pairing condition and in the presence of flux $N_\phi$. The $L=S=0$ state at $N_\phi=1$ is the bosonic NASS state, the state with
$L=0$, $S=2$ at $N_\phi=2$ is the bosonic MR state.}

\label{N4Nflux1_2_3}
\end{table}

The idea is now to extend the trial states eq.~(\ref{symI_II}) to
the spin-full case, by including factors of type $\psi_{\rm B}$ separately
in both group $I$ and group $I\!I$. All
states generated in this way satisfy the pairing property. They
constitute a subset of all paired states at $N_\phi=N_\phi^{\rm MR}+1$
as listed in table~\ref{N4Nflux1_2_3}.

% sub-problem: N/2 + N/2 particles in 2 orbitals

We first analyze the symmetric product of the states $\psi_{\rm B}$
for groups $I$ and $I\!I$. In $SU(4)$ group theory
\be
Z_{\rm B}^{I,I\!I}
=  [N/2 \, 0 \, 0] \otimes_{\rm sym} [N/2 \, 0\, 0]  
=  \sum_{l=0}^{N/4} [N-4l \, 2l \, 0]  \ . \nn
\ee
For $N=4$, the $l=0$ contribution has $(L,S)=(2,2)$, $(1,1)$ and
$(0,0)$, totaling 35 states, while the $l=1$ term comprises
$(L,S)=(2,0)$, $(1,1)$, $(0,2)$, $(0,0)$, totaling 20 states.
For general $N$, $l$, the representation $[N-4l \, 2l \, 0]$
contains fully polarized states ($S=N/2$) at $L=N/2-2l$. 
Note that if we were to fully symmetrize over all $N$ particle
coordinates, only the $l=0$ term would survive, reducing the
construction to the states $|\psi_{\rm B}^{K}\rangle$ describing 
the iqH skyrmion.
%
%\begin{table}
%
%\begin{tabular}{|c|c|c|c|}
%\hline
%        & S=0 & S=1 & S=2 \\
%\hline
%L=0 & 1 & 0 & 1 \\
%\hline
%L=1 & 0 & 2 & 0 \\
%\hline
%L=2 & 1 & 0 & 1 \\
%\hline
%\end{tabular}
%\hspace{2mm}
%\begin{tabular}{|c|c|c|c|c|c|}
%\hline
%        & S=0 & S=1 & S=2 & S=3 & S=4 \\
%\hline
%L=0 & 3 & 0 & 2 & 0 & 1 \\
%\hline
%L=1 & 0 & 4 & 1 & 2 & 0 \\
%\hline
%L=2 & 2 & 1 & 4 & 1 & 1 \\
%\hline
%L=3 & 0 & 2 & 1 & 2 & 0 \\
%\hline
%L=4 & 1 & 0 & 1 & 0 & 1 \\
%\hline
%\end{tabular}
%
%\caption{Multiplicities of $(L,S)$ multiplets captured by the
%expression ${\rm Symm}_{I,I\!I}[\psi_I\psi_{I\!I}]$ as detailed in the
%text, for $N_\phi=N_\phi^{\rm MR}+1$ and $N=4$ (left panel) and
%$N=8$ (right panel).}
%
%\label{N4_8flux3_7}
%\end{table}

% LLL lift and symmetrization

However we first perform what we call the Lowest-Landau-Level lift 
%(amounting to the insertion of $\nu=1/2$ Laughlin factors in both groups),
of $\psi^{I,I\!I}_B$
\[
\psi^{I,I\!I}_B \rightarrow \psi^{I,I\!I}_B \left(\{x_{i \in I \cup I \! I} \}\right)\; \psi^L_{I} \left(\{x_{j \in I} \}\right) \; \psi^L_{I \! I}\left(\{x_{k \in I \! I} \}\right),
\]
where the sets of coordinates $\{ x_i \}$ contain both the up-spins $z_i$ and down-spins $w_i$. 
The polynomial $\psi_B$ is then symmetric with respect to exchanging $z_i$ and $z_j$ if $i,j$ are in the same
group, and similarly for the down spin coordinates; we require no further symmetries.
If we only then symmetrize over
all $N$ particles, we keep a much bigger set of spin-full states
satisfying the pairing condition.
%See table \ref{N4_8flux3_7} for $N=4$ and $N=8$.
In general states obtained for 2 orbitals lift to
independent states at $N_{\phi}=N-1$. One exception is
$(L,S)=(0,0)$ at $N=4$, where the LLL-lifts from the $l=0,1$ 
multiplets coincide.

We now argue that the states that survive after the symmetrization step with $l>0$ can be
viewed as charge $q$ CST separated by a distance set by $l$, where 
$l=N/4$ corresponds to the situation that two $q$ CST sit on opposite poles of the sphere. 
This is most easily seen by focussing on the fully polarized
($S=N/2$) states, where the expressions can be compared to explicit 
formulas describing spin-less charge $q$ quasiholes. The
states (in disc geometry) for $N$ paired, 
spin-polarized bosons at $N_\phi=N-1$ 
are obtained by expanding the 2-quasihole wavefunction
\be
\begin{array}[t]{c} {\rm Symm} \\ {\small I,I\!I} \end{array}  \prod_{i \in I}(\eta_1-z_i) \!\! \prod_{i<j \in I}(z_i-z_j)^2 \!
\prod_{k \in I\!I}(\eta_2-z_k) \!\! \prod_{k<l \in I\!I} \!\! (z_k-z_l)^2 \nn
\ee
on a basis of symmetric polynomials in $\eta_1$, $\eta_2$, where the
powers of the $\eta_s$ indicate the location of the two quasiholes.
On the sphere, the resulting angular momenta are 
(for $N$ a multiple of $4$)
$L=N/2, N/2-2, \ldots, 0$. To leading order in $1/N$, the state 
at $L=0$ corresponds to $\eta_1^{N/2}\eta_2^0 + \eta_1^0\eta_2^{N/2}$ 
indicating that indeed the two quasiholes are on opposite sides of 
the sphere. For the spin-full case, one similarly finds that 
2-CST states with $L\ll N/2$ correspond to well-separated CSTs.

For working towards explicit expressions, it is
most convenient to perform the LLL lift and subsequent
symmetrization in second quantization. Within each group, the LLL 
lift amounts to an embedding of a state defined on 2 orbitals to one 
on $N$ orbitals, with coefficients set by the expansion of the
corresponding Laughlin factor. It has the important property that both 
$L$ and $S$ quantum numbers are preserved. For the simple example of 
the 2-particle, polarized $\nu=1/2$ Laughlin state (corresponding
to one of the groups $I$, $I\!I$ for $N=4$ particles and $N_{\phi}=3$ flux quanta) 
the LLL-lift takes the form
\beq
 | \! \up_2, 0 \rangle &\rightarrow& 
  2\sqrt{6} | \! \up, 0 ,\up , 0 \rangle - 4 |0, \up_2 ,0, 0\rangle
\nonumber \\
\label{LLL-4}
 | \! \up,\up\rangle &\rightarrow& 
  6 | \! \up,0 , 0,\up \rangle - 2 |0, \up, \up, 0\rangle
 \\
 |0,\up_2\rangle &\rightarrow&
  2\sqrt{6} |0,\up,0 ,\up\rangle - 4 |0,0, \up_2 ,0\rangle \nonumber \ . 
\eeq
We will work out the second line above as an example. In first quantization
this state corresponds to the lift of $(z_1 + z_2)$:
\beq
\label{LLL-1q}
(z_1 + z_2) (z_1 - z_2)^2 = (z_1^3 + z_2 ^3) - (z_1 z_2^2 + z_1^2 z_2),
\eeq
where we have expanded on a basis of symmetric monomials. When going from
first to second quantization on the sphere, particles in orbital
$l=0,\dots , N_\phi$ obtain an additional factor $\sqrt{l!}\sqrt{(N_\phi -l)!}$, giving
\beq
(z_1^3 + z_2 ^3) &\rightarrow& \sqrt{6}\sqrt{6}| \! \uparrow,0,0,\uparrow\rangle \nonumber \\
(z_1 z_2^2 + z_1^2 z_2) &\rightarrow& \sqrt{2}\sqrt{2}|0,\up, \up ,0\rangle \nonumber,
\eeq
which leads to (\ref{LLL-4}) after we plug in the relative coefficients found in the expansion (\ref{LLL-1q}).
The symmetrization over groups $I$ and $I\!I$ is easily done through the step
\beq
\lefteqn{ 
|\ldots, m_I, \ldots \rangle \times_s |\ldots, m_{I\!I}, \ldots \rangle
\rightarrow }
\nonumber \\[2mm]
&& \sqrt{ \frac{(m_I+m_{I\!I}) !}{m_I ! \; m_{I\!I} !} }
   |\ldots, m_I + m_{I\!I}, \ldots \rangle ,
\eeq
where $m_I$, $m_{I\!I}$ indicate the occupation number of a given orbital
(including its spin-label).

As an explicit example, we present the two $L=S=1$ 
states for $N=4$ particles with $N_\phi=3$. In the first step we identify the
$L=S=1$ highest weight states within the two distinct $SU(4)$ multiplets
with Dynkin labels $[4\,0\,0]$ and $[0\,2\,0]$, 
\beq
&& \psi_{\rm B}^{[4\,0\,0]} \propto [\sqrt{2} | \! \uparrow, \uparrow \Uparrow \Downarrow \rangle
         + | \! \uparrow, \downarrow \Uparrow_2 \rangle
         -3 | \! \downarrow, \uparrow \Uparrow_2 \rangle] + [I \leftrightarrow I\!I]
\nonumber \\
&& \psi_{\rm B}^{[0\,2\,0]} \propto [-\sqrt{2}| \! \uparrow, \uparrow \Uparrow \Downarrow \rangle
          +2 | \! \uparrow, \downarrow \Uparrow_2 \rangle] + [I \leftrightarrow I\!I] 
\eeq
where we use single (double) arrows for indicating the particles in group $I$ ($I\!I$).
In the second step we perform the LLL-lift and then symmetrize, leading to
%(more details on how we do this??)
\beq
\lefteqn{\psi^{[4\,0\,0]} \propto 5 |0,\downarrow, \uparrow_3,0 \rangle
- 3\sqrt{2} |0,\uparrow\downarrow, \uparrow,\uparrow \rangle}
\nonumber \\ &&
\;\; + 7 |0,\uparrow_2, \downarrow,\uparrow \rangle
- {5 \over 3}\sqrt{3}|0,\uparrow, \uparrow_2\downarrow,0 \rangle
- |0,\uparrow_2, \uparrow,\downarrow \rangle
\nonumber \\ &&
\;\; - \sqrt{3}| \! \downarrow,0, \uparrow_2,\uparrow \rangle
+ 3 | \! \downarrow,\uparrow,0, \uparrow_2\rangle
- \sqrt{6} | \! \uparrow,0, \uparrow\downarrow,\uparrow \rangle
\nonumber \\ &&
\;\; + 3\sqrt{3} | \! \uparrow,0, \uparrow_2,\downarrow \rangle
+ 3 | \! \uparrow,\downarrow,0, \uparrow_2 \rangle
- 3 \sqrt{2} | \! \uparrow,\uparrow,0,\uparrow\downarrow \rangle
\nonumber
\eeq
\beq
\lefteqn{\psi^{[0\,2\,0]} \propto 
4 |0,\downarrow, \uparrow_3,0 \rangle
- {3 \over 2}\sqrt{2} |0,\uparrow\downarrow, \uparrow,\uparrow \rangle}
\nonumber \\ &&
\;\; + 2 |0,\uparrow_2, \downarrow,\uparrow \rangle
- {4 \over 3}\sqrt{3}|0,\uparrow, \uparrow_2\downarrow,0 \rangle
+ |0,\uparrow_2, \uparrow,\downarrow \rangle
\nonumber \\ &&
\;\; - 2\sqrt{3}| \! \downarrow,0, \uparrow_2,\uparrow \rangle
+ 6 | \! \downarrow,\uparrow,0, \uparrow_2\rangle
+ \sqrt{6} | \! \uparrow,0, \uparrow\downarrow,\uparrow \rangle
\nonumber \\ &&
\;\; - 3 | \! \uparrow,\downarrow,0, \uparrow_2 \rangle
- {3 \over 2} \sqrt{2} | \! \uparrow,\uparrow,0,\uparrow\downarrow \rangle \ .
\label{L1S1states}
\eeq
We remark, while the construction correctly reproduces two linearly independent
paired states at $L=S=1$, the algebraic structure is not very transparent. For 
one thing, the $SU(4)$ symmetry is lost in the LLL-lift. In addition, the two states 
eq.~(\ref{L1S1states}) are not  orthogonal.
 
The $(L,S)$ states constructed here can directly be compared to the numerical 
ground states found in \cite{WMSC}. In particular, where \cite{WMSC} finds $L=0,1$ 
groundstates for $N=12$ at $S=4,5,6$ we expect good overlaps with the states 
presented here.

% semiclassical limit and spin-texture
The bosonic parts of the wavefunctions simplify considerably in the large-$N$ limit:
the leading polarized states of the $[N\,0\,0]$ and $[0\, N/2 \, 0]$ irreps are
\be
 \psi_{\rm B}^{[N\,0\,0]} \rightarrow | 0,\uparrow_{N/2} \Uparrow_{N/2} \rangle, \quad
 \psi_{\rm B}^{[0\, N/2\, 0]} \rightarrow | \! \uparrow_{N/2},\Uparrow_{N/2} \rangle . \nn
\ee

We can then consider states with $K$ overturned spins as in eq.~(\ref{Kturned}), 
separately in groups $I$ and $I\!I$. This leads to simple expressions for the
situation where the spin-textures in groups $I$, $I\!I$ have sizes
$\lambda_I$, $\lambda_{I\!I}$. Starting from $\psi_{\rm B}^{[0\, N/2\, 0]}$,
with group $I$, $I\!I$ textures on opposite sides of the 2-orbital subspace, 
and then taking these expressions through the LLL lift and symmetrization, 
leads to 2-CST configurations where the charge $q$ CSTs sit on opposite
sides of the sphere. This expression, expanded in powers of the sizes $\lambda_I$ and $\lambda_{I\!I}$, 
symbolically reads
\be 
\label{expans}
\sum_{K_I,K_{I \! I}} \lambda_I^{K_I} \lambda_{I\!I}^{K_{I \! I}}
\left( \begin{array}[c]{c} {\rm Symm} \\ {\small I,I\!I} \end{array} \right)
\left( {\rm LLL-lift} \right)
 | \! \uparrow_{N/2-K_I} \Downarrow_{K_{I\!I}}, 
  \downarrow_{K_I}   \Uparrow_{N/2-K_{I\!I}}\rangle \ . 
\ee
From here on we will label our textures as CST[$w_I$,$w_{I\!I}$], where $w_I$, $w_{I\!I}$ 
are the winding numbers (with respect to a given location on the sphere) of the skyrmions 
that would appear in group $I$, $I\!I$ if  the symmetrization step in our construction were not performed. 
The $SU(4)$ label $[N \, 0 \, 0]$ corresponds to the CST$[1,1]$ and $[0\, N/2 \, 0]$ gives
two spatially separated CST$[1,0]$.

% spin textures

\begin{figure}[h!]
\begin{center}
\includegraphics[width= .3\textwidth]{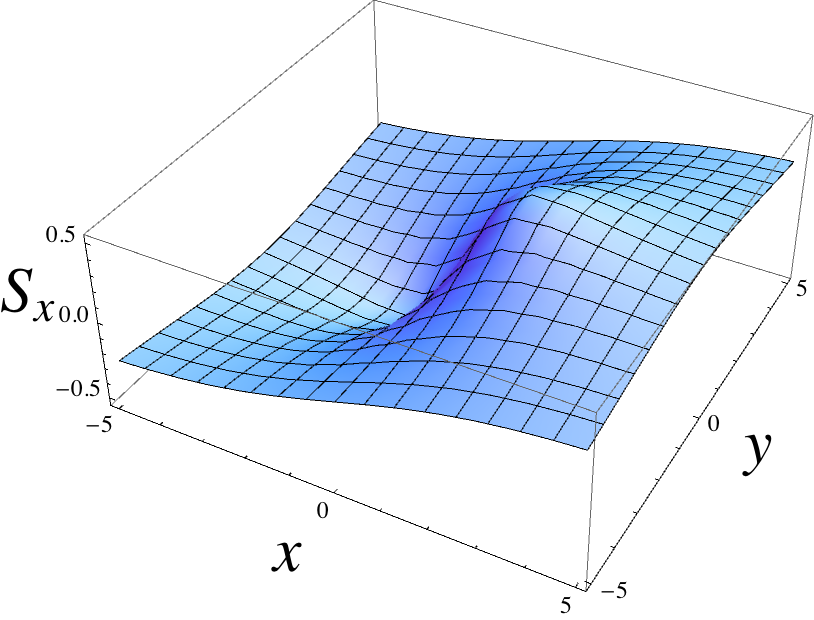}
\hspace*{2mm}
\includegraphics[width= .3\textwidth]{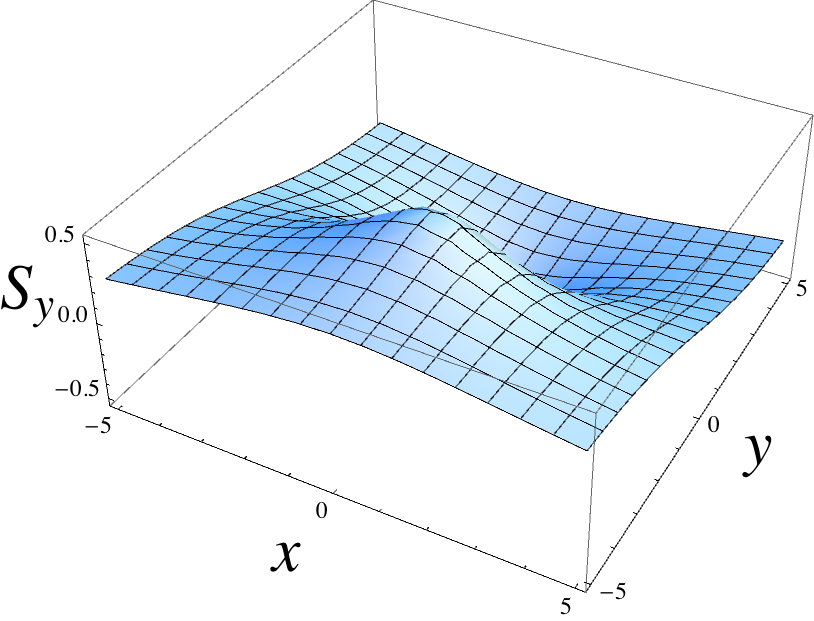}
\hspace*{2mm}
\includegraphics[width= .3\textwidth]{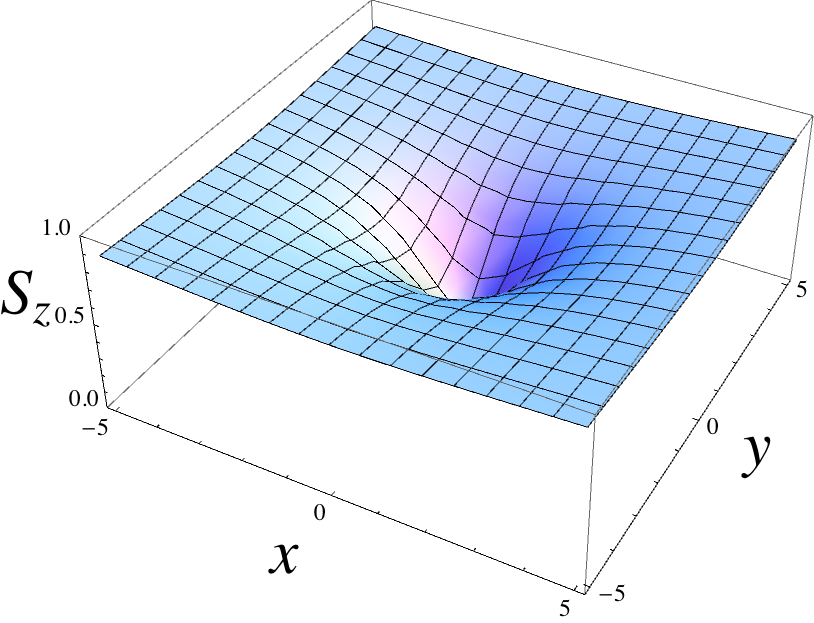}
\caption{Spin components $S_x$, $S_y$ and $S_z$ for configuration with single CST$[1,0]$ 
at the origin and a quasihole at infinity, after stereographic projection.} 
\label{fig:N8texture}
\end{center}
\end{figure}

For a given quantum state we measure the components of the spin field by acting with
 \be
{\bf S}(z,\zb) = \sum_{l,l'}
\left( a^\dg _{l,\alpha} {\bf \sigma}_{\alpha\beta}   a_{l',\beta} \right)  \phi_l^\dg (\zb)\; \phi_{l'} (z) \nn
 \ee
% \beq
% S_x (z, \zb) &=& \sum_{l,l'} \left( a^\dg _{l \up} a_{l' \down} + a^\dg _{l \down} a_{l' \up} \right) \% phi_l^\dg (\zb)\; \phi_{l'} (z), \nn \\
% S_y (z, \zb) &=& \sum_{l,l'} \left( -i \; a^\dg _{l \up} a_{l' \down} + i\; a^\dg _{l \down} a_{l' \up} \right) \phi_l^\dg (\zb)\; \phi_{l'} (z), \nn \\
% S_z (z, \zb) &=& \sum_{l,l'} \left( a^\dg _{l \up} a_{l' \up} - a^\dg _{l \down} a_{l' \down} \right) \phi_l^\dg (\zb)\; \phi_{l'} (z),
%\eeq
where the $\phi_l(z)$ are the single particle wave functions, which depend on the geometry and our normalization is such that a polarized system has $S_z=1$. In figure \ref{fig:N8texture} we plot the expectation value of the spin vector for a configuration 
with $N=8$, $\lambda_I=1.0$,  $\lambda_{I\!I}=0.0$. 
%After stereographic projection this corresponds to a CST$[1,0]$ at the origin and a single quasi-hole at infinity.

%For $N$ odd the MR state has unpaired fermion, meaning that a collection of $n$ fundamental quasiholes fuse into the so-called $\psi$-channel.
In a Conformal Field Theory (CFT) based description \cite{MR}, the quasihole operator comes with an Ising $\sigma$-field. For $N$ even, a collection of these $\sigma$-fields will fuse to the identity operator, whereas they fuse to the $\psi$ sector for $N$ odd. This is due to the fact
that the electron operator carries a $\psi$ field, so performing the contractions of an odd number of electron operators within the CFT correlator will always
leave one $\psi$ field: this one has to pair with the fusion product of all the $\sigma$-fields in order for the total correlator to fuse to the identity.

It was shown in \cite{PH} that the density profile
of the system on the sphere after the fusion of two charge $q$ quasiholes to a charge $2q$ quasihole differs
between the two cases: in the case $N$ even the density drops to zero at the location of the $2q$ quasihole, whereas the
density drop for $N$ odd is wider and less deep.

This has consequences for the possible spin textures that may arise as a result of fusing elementary CST$[1,0]$.
Our construction recovers the polarized quasihole states in the limit $\lambda \rightarrow 0$. This means we expect the
density for up-particles to vanish in the core of the fusion product of two CST$[1,0]$ for $N$ even, but not for
$N$ odd.

This becomes rather obvious in the two-layer construction we have been using throughout this paper. For
$N$ even, the particles can be divided equally into two groups. The $K_I=K_{I\! I}=0$ term appearing in 
the expansion of two CSTs analogous to (\ref{expans}), but now for both CSTs at the same position
is then
\[
|0, \up_{N/2} \rangle \textrm{ for group $I$},\;\;\;|0, \Up_{N/2} \rangle \textrm{ for group $I\! I$}.
\]
However for $N$ odd we have to divide the particles unequally among the two groups. The division that requires the
least amount of total flux is $(N+1)/2$, $(N-1)/2$. The $N_\phi$ is equal for both groups, and is at least $N-1$: this is the
highest power for a single particle appearing in the expansion of Laughlin factor for the group containing $(N+1)/2$ particles
\footnote{Note one can also divide the particles 
unequally in the case $N$ even. This leads to states with quasiholes: consider the division $N/2+1$, $N/2-1$. The particles in the first
group require at least $N_\phi=N$, which means the particles in the second group have four excess fluxes.}. Note that the $N_\phi^{\mathrm{MR}}=N-2$,
so that paired states with an odd number of particles will always have quasiholes present.

This extra flux in the system gives
two extra orbitals to the particles in the smaller group, whereas the particles in the larger group have no additional orbitals.
The first term in the expansion now becomes
\[
|0, 0, \up_{(N-1)/2} \rangle \textrm{ for group $I$},\;\;\;| \! \Up_{(N+1)/2} \rangle \textrm{ for group $I\! I$}.
\]
Even without performing the whole calculation one can already see that the density will not vanish since the particles
in group $I\! I$ are spread homogeneously over the sphere. The natural texture
\be 
\sum_{K} \lambda^{K}
\left( \begin{array}[c]{c} {\rm Symm} \\ {\small I,I\!I} \end{array} \right)
\left( {\rm LLL-lift} \right)
 | \! \downarrow_{K},0, \uparrow_{(N-1)/2-K}\rangle  | \! \Uparrow_{(N+1)/2}\rangle, 
\ee
has winding number 2 for group $I$ and 0 for group $I \! I$. Therefore we argue that 
the simplest possible charge $2q$
configuration has winding indices $[2,0]$ for $N$ odd. 

%In general
%$
%w_I - w_{I\! I} \equiv 2N \mathrm{mod} \; 4 .
%$
%The fact that the spin textures correlate with the fusion channel of fundamental
%quasiholes offers possibilities for the read-out of topological qubits. The encoding of the fusion channel
%in the division of the particles into the two groups turns out to be general. In particular, we found
%that the possible ways to divide the particles are one-to one with classes of possible paths through the Bratelli diagram
%of the underlying CFT. This claim is the object of further study \cite{future}.

It is natural to ask the question what spin textures arise when multiple CST$[1,0]$ are fused together. Depending on the
path one takes through the Bratelli diagram of the CFT of the underlying quasi holes, multiple options are possible.
In \cite{RR} the authors define the number of unpaired fermions $F$ in order to write down wave functions
in different fusion channels. For a general number of extra fluxes $\Delta N_\phi$ above the ground state $0 \leq F \leq \Delta N_\phi$ and
$F=N \; \mathrm{mod} \; 2$. We obtain the same wave functions by dividing the particles as $(N+F)/2$ in group $I$ and $(N-F)/2$ in group $I\! I$.
The associated spin textures satisfy the identity $|w_I - w_{I\! I}| = 2F$. For $\Delta N_\phi=1$ and $F=0\;(1)$ this corresponds to the even (odd) discussion
in the above paragraph. The relation between the formulation of wave functions
in terms of this $F$, the multiple group construction and those obtained through CFT will be further presented in an upcoming work \cite{future}.

\begin{figure}[h!]
\begin{center}
\includegraphics[width= .35\textwidth]{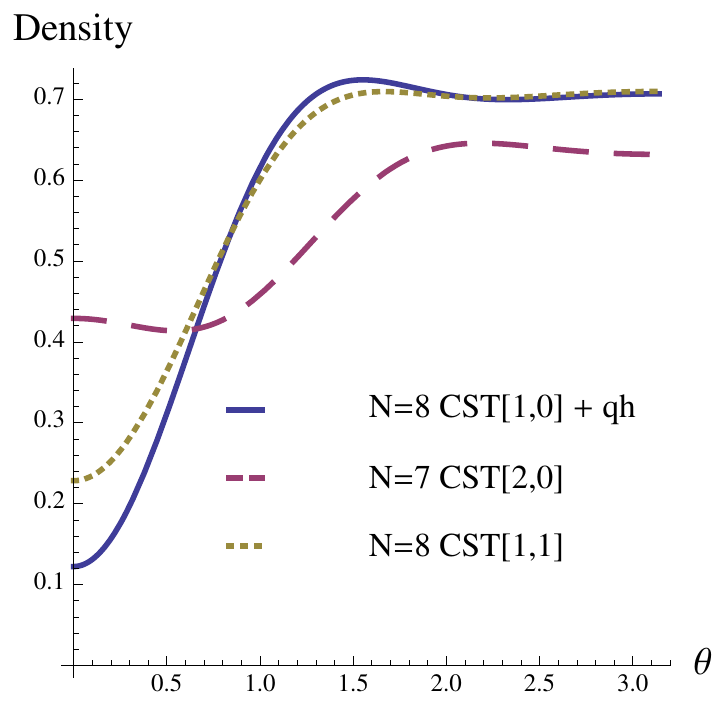}
\hspace{3mm}
\includegraphics[width= .35\textwidth]{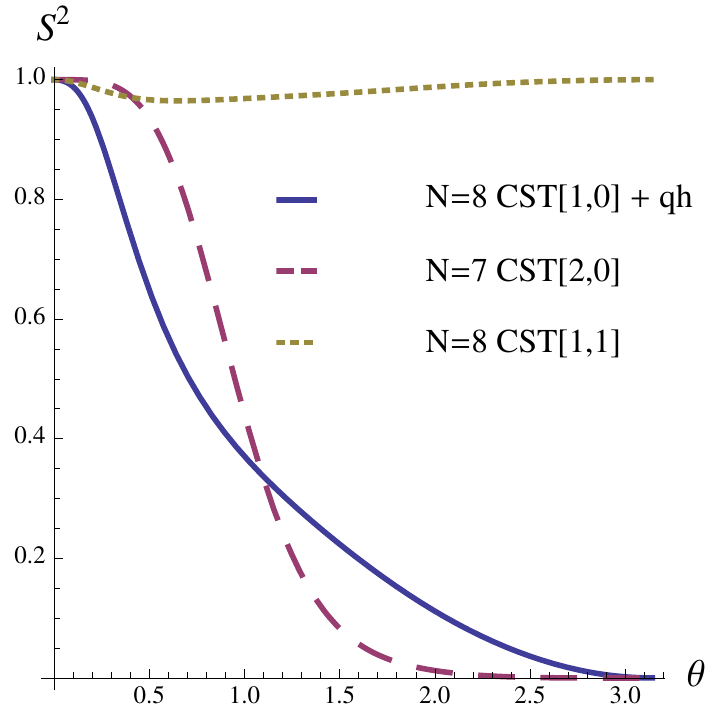}

\includegraphics[width= .35\textwidth]{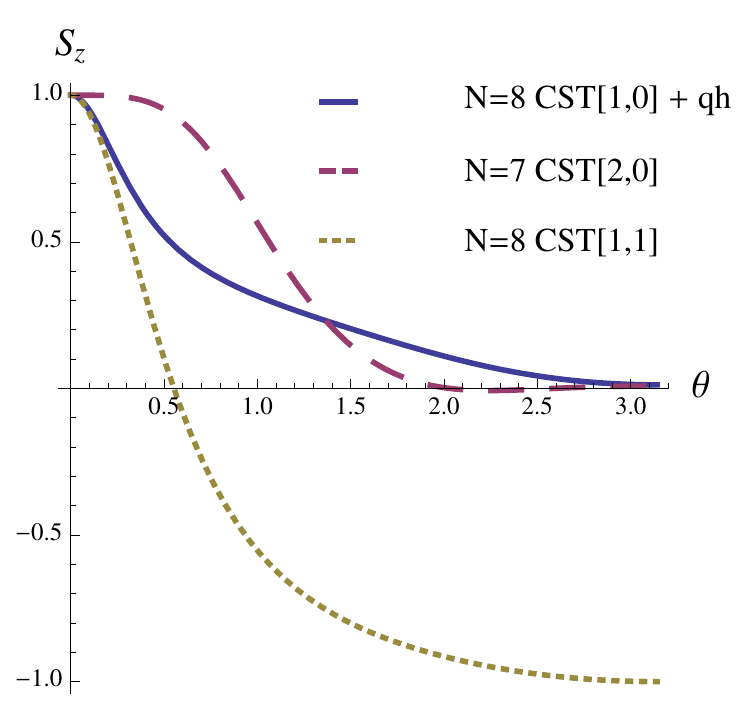}
\hspace{3mm}
\includegraphics[width= .35\textwidth]{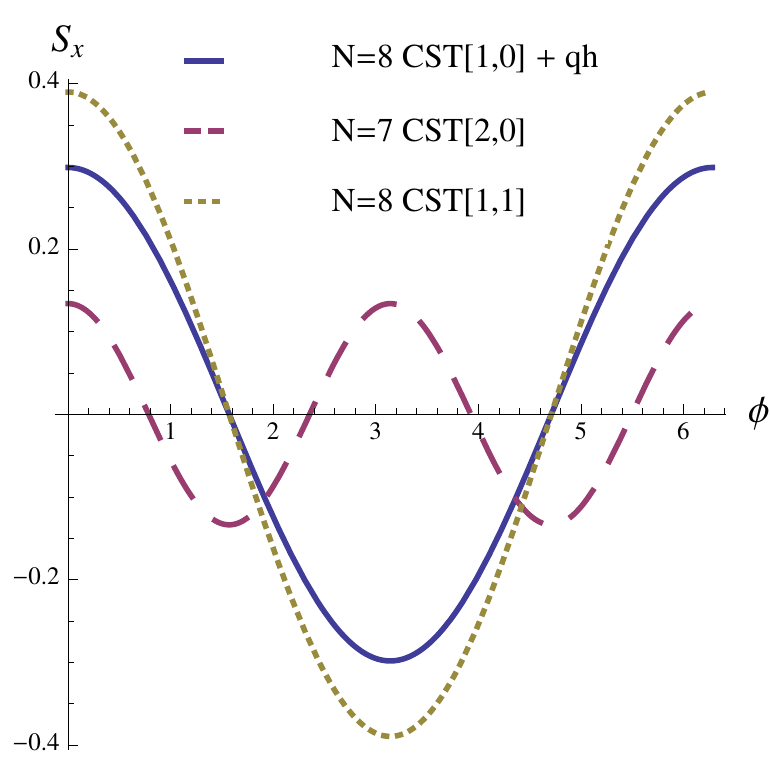}
\caption{Density, $S^2$ and $S_z$ of a $N=8$ CST$[1,0]$/quasihole pair, an $N=7$ CST$[2,0]$ and an $N=8$ skyrmion (CST$[1,1]$) at $\lambda=2.0$ as a function of the polar angle $\theta$ 
at azimuthal angle $\phi=0$ and $S_x$ for the same systems as a function of $\phi$ at 
$\theta=\frac{\pi}{2}$. The spin textures are centered at $\theta=\pi$, for the $N=8$ CST$[1,0]$ 
there is a quasihole at $\theta=0$.} 
\label{fig:2dplots}
\end{center}
\end{figure}

We have studied three representative cases in detail: a skyrmion CST$[1,1]$ for $N=8$, a separated  CST$[1,0]$/quasihole pair for $N=8$ and a single CST$[2,0]$ for $N=7$. 
The results are in figure \ref{fig:2dplots}. We have chosen these cases for the following reasons. The CST$[1,0]$ is the fundamental charge $q$ spin texture,
associated with the $\sigma$-field in the Ising CFT. The skyrmion CST$[1,1]$ is given because it shows that our 
construction includes the results of earlier studies \cite{MFB}. It is also the fusion product, following the discussion above, of two elementary CST$[1,0]$ in the
trivial ($N$ even) fusion channel. The CST$[2,0]$ is the fusion product of two CST$[1,0]$ when
the overall fusion channel (in CFT language) is $\psi$ or alternatively stated, when the number of particles $N$ is odd. 

Two observations about the behaviour of these textures are in place. First of all we see that the the CST$[2,0]$ has winding number 2 when the azimuthal angle 
runs from $0$ to $2 \pi$. Furthermore, the (expectation value of the) length of the spin vector vanishes in the core of the CSTs of type $[1,0]$ and $[2,0]$.
For $N$ large the latter effect seems to hold for all CST$[n,0]$. This behavior closely mimicks that of the `polar core vortex' appearing in rotating spin-1 Bose-Einstein condensates \cite{MKM}. The observation that the MR state carries an effective spin-1 field due to the pairing of the electrons has been made in earlier studies \cite{YR,DHN}. The BEC polar core vortices have the following mean field spin vector expectation value
\be
{\bf S}(r,\phi) = (\sqrt{2\rho(1-\rho)} \cos n \phi,\;- \sqrt{2\rho(1-\rho)} \sin n \phi,\; \rho)\; , \nn
\ee
with $\rho(r)$ equal to $0$ at the origin and approaching $1$ at infinity.
The integrated Pontryagin density for these textures equals $Q_{\rm top} = \frac{n}{4}$. Numerical values for 
our CST$[1,0]$ textures approach $\frac{1}{4}$ for $\lambda \gg 1$. For a general 
texture CST$[w_I,w_{I\! I}]$ the integrated Pontryagin density is no longer a topological index 
in the usual sense (the target space manifold is $\mathbb{R}^3$ instead of $S^2$, so the integral does not have to be an integer). Also, the relation between electric and topological charge densities, $\rho_{\rm elec} = \nu e \rho_{\rm top}$, valid in Abelian quantum Hall states, takes a different form in general non-Abelian states, of which the MR state is a prototypical example. 

Having completed this work, we have been informed of related but unpublished work of 
I.~Dimov and C.~Nayak. They considered 1st quantized expressions for CST based
on the `Pfaffian' expression of the MR state and examined the associated spin textures, 
observing as we did the vanishing of the spin vector in the core of the CST. 
It seems clear that for the purpose of numerically generating finite-$N$ trial states the procedure proposed here (which avoids cumbersome expansions of expressions in 1st quantization) is particularly efficient.

\vskip 4mm

We thank Chetan Nayak for explaining the unpublished work
and Nigel Cooper, Steve Simon, Sander Bais and Jan de Boer for discussions. J.R. is financially 
supported by FOM. L.H. and K.S. acknowledge hospitality by the 
Aspen Physics Centre and by NORDITA, where part of this work was done.
L.H. acknowledges support from ICAM-I2CAM, 1 Shields Avenue, Davis, 
CA 95616 (NSF grant DMR-0844115).
\\

\end{document}